\documentclass[prd,showpacs,preprintnumbers,nofootinbib,aps]{revtex4}
\usepackage{subeqnarray}
\usepackage{epsfig,amsmath,amssymb}
\usepackage{mathrsfs}
\usepackage[usenames,dvipsnames]{color}
\usepackage[pagebackref=true, colorlinks=true]{hyperref}
\definecolor{redish}{rgb}{0.7,0.2,0.0}  
\definecolor{bluish}{rgb}{0.2,0.5,0.8}
\hypersetup{linkcolor=redish,          
                  citecolor=blue,        
                  filecolor=magenta,      
                  urlcolor=bluish}          
\numberwithin{equation}{section}
\DeclareFontFamily{U}{rsfs}{}         
\DeclareFontShape{U}{rsfs}{m}{n}{<5> rsfs5 <6><7> rsfs7          %
  <8><9><10><10.95><12><14.4><17.28><20.74><24.88> rsfs10}{}     %
\DeclareMathAlphabet{\mathfs}{U}{rsfs}{m}{n}                     %
\newcommand{\mfs}[1]{\mathfs {#1}}                               %

\newcommand{\ba}{\nopagebreak[3]\begin{eqnarray}}
\newcommand{\ea}{\end{eqnarray}}
\newcommand{\bii}{\begin{itemize}}
\newcommand{\eii}{\end{itemize}}
\newcommand{\nn}{\nonumber}

\newcommand{\sA}{{\mfs A}}
\newcommand{\sL}{{\mfs L}}
\newcommand{\f}{\frac}
\def \p{\partial}
\def \d{\delta}
\def \b{\beta}
\def \l{\ell}
\def \g{\gamma}
\def \lp{\ell_p}
\def \j{\sqrt{j(j+1)}}

\def \lm{\lambda}

\def \s{\sigma}

\begin{document}
\title{Stability of Quantum Isolated Horizon : A Local Observer's view}
\author{Abhishek Majhi}%
\email{abhishek.majhi@saha.ac.in}
\affiliation{Saha Institute of Nuclear Physics,\\Kolkata 700064, India}%
\pacs{04.70.-s, 04.70.Dy}

\begin{abstract}

It is shown that a Quantum Isolated Horizon(QIH), {\it as observed by a local observer}, is locally \textit{unstable} as a thermodynamic system. The result is derived in two different ways. Firstly, the specific heat of the QIH is shown to be negative definite through a quantum statistical analysis. Then, it is shown, in the thermal holographic approach, that the canonical partition function of the QIH diverges under Gaussian thermal fluctuations of such energy spectrum, implying local instability of such a QIH as a thermodynamic system. \end{abstract}
\maketitle
\section{Introduction}
In Loop Quantum Gravity(LQG), the quantum description of black hole is captured by the notion of Quantum Isolated Horizon (QIH). The effective description of a QIH is given by a three dimensional SU(2) Chern Simons (CS) theory coupled to point like sources on punctures made by the edges of spin network describing
the bulk quantum geometry \cite{qg,abck,km4}. The punctures are associated with SU(2) spins and designated by quantum numbers $j,m$. The microstates of the CS theory give rise to the {\it microcanonical} entropy of the QIH \cite{qg,abck,km4,mpre}.  For a complete understanding of the thermodynamical aspects of a black hole, besides finding the entropy, we need to find the energy spectrum of a QIH i.e. energy quantization for a black hole, which is yet to be done.

\par
Recently, there have been proposal of a quantum energy spectrum of QIH \cite{GP} based on semi-classical approximations and arguments from classical Schwarzschild black hole, which are used as inputs from outside the {\it pure} quantum theory, so as to avoid a true quantization of the black hole energy. The proposed energy spectrum of the QIH in \cite{GP} is given by 
\ba
\widehat E~|j_1, \cdots ,j_N\rangle=\f{1}{8\pi\l}~\widehat A~|j_1, \cdots ,j_N\rangle\label{ea}
\ea
 where $\widehat E$ and $\widehat A$ are the Hamiltonian and area operators for the QIH respectively.                The proposal of the energy spectrum in eq.(\ref{ea}) follows form the semiclassical relation $E=A/8\pi\l$, elaborated in \cite{GP2}, where $E=$ the classical energy of the black hole , $A=$ the classical area of the horizon and $\l$ is a {\it constant} length scale characterizing the proper distance of a class of stationary local observers from the horizon, introduced in \cite{GP2}. Similar expression for local energy $(E=A/8\pi\l)$ also appears in \cite{bia,smo} with explanations on different grounds \footnote{As far as this length scale $\l$ is concerned there is a conflict between \cite{GP,GP2} and \cite{bia}. According to \cite{GP,GP2}, $\l\sim\lp$. On the other hand, according to \cite{bia}, $\l\gg\lp$, $\lp$  being the Planck length.}. Since only \cite{GP} discusses the thermodynamic stability of the QIH and thus, more closely related to the subject matter of this paper, we shall only refer to \cite{GP} at the relevant places.

\par
Here, we present a detailed stability analysis of a QIH having the energy spectrum as in eq.(\ref{ea}), to argue that such a spectrum in fact leads to the local thermodynamic {\it instability} of the QIH. The stability analysis is carried out using two methods -- one that is similar to the one followed in \cite{GP} and the other which is a completely independent approach, namely the thermal holographic method introduced in \cite{pm1,ampm}. Both methods lead to the same conclusion which is the key result of this paper: {\it  An uncharged,  non-rotating QIH, as observed by the local observers discussed in \cite{GP,GP2}, is locally unstable as a thermodynamic system.} An outline of the contents of this paper is briefly stated as follows.

\par
Section (\ref{summe}) deals with a short review on the {\it model independent} derivation of the {\it microcanonical} entropy of a generic QIH from the microstates of the QIH Hilbert space \cite{mpre}. In section (\ref{es}), the importance of this particular energy spectrum (eq.(\ref{ea})) proposed in \cite{GP,GP2}, is discussed in the context of its consequences as far as current research is concerned \cite{bia,smo}.  In section (\ref{tqih}) the thermodynamic stability of the QIH is analyzed in complete detail. Effects of both {\it quantum} and {\it thermal} fluctuations are incorporated. First, in subsection (\ref{eqssa}) thermodynamic quantities associated with the QIH are calculated using an explicit quantum statistical formulation. The specific heat is found to be negative definite showing the QIH to be locally {\it unstable} as a thermodynamic system.  In subsection (\ref{gtf}), the effect of thermal fluctuations on the stability of the QIH is investigated in the thermal holographic approach \cite{pm1,ampm}, once again resulting in the local {\it instability} of the QIH.  Finally, we conclude with a discussion  in section (\ref{d}).


\section{SU(2) Microstates and Microcanonical Entropy}\label{summe}

In this section we present a short review of  the derivation of the {\it microcanonical} entropy$(S_{MC})$ of a generic QIH shown in \cite{mpre} where $S_{MC}$ is obtained from the  fundamental statistical mechanical analysis of the quantum degrees of freedom arising from the Hilbert space associated with the QIH. The crucial steps are debriefed as follows :

\par
{\it The Microstates of a QIH }: The Hilbert space associated with a QIH, as mentioned earlier, is that of the SU(2) CS theory coupled to point like sources (punctures  made by bulk spin network on the Isolated Horizon) on the QIH.  The counting of the SU(2) microstates have been done Kaul-Majumdar in \cite{km1} using the relation between the Hilbert space of the SU(2) CS theory on the boundary(QIH) with the space of conformal blocks of the Wess-Zumino model  on the boundary 2-sphere. The number of microstates for a spin sequence $\{j_1,j_2, .....,j_N\}$  is given by
\ba
d[\{j_l\}]&=&\f{2}{k+2}\sum_{q=0}^{k/2}\frac{\prod_{l=1}^{N}\sin \left[\frac{(2j_l+1)(2q+1)\pi}{k+2}\right]}{\left\{\sin\left[\frac{(2q+1)\pi}{k+2}\right]\right\}^{N-2}}\nn
\ea
where $j_l$ is the spin associated with the $l$th puncture, $k$ is the level of the SU(2) CS theory and $N$ is the total number of punctures.  Alternatively, we can switch over to the spin configuration basis. The number of SU(2) microstates for a spin configuration $\left\{s_j\right\}$ is given by
\ba
d[\left\{s_j\right\}]=\f{2}{k+2}\frac{\left(\sum_{j}s_j\right)!}{\prod_{j}s_j!}~~\sum^{k+1}_{a=1}\sin^2\frac{a\pi}{k+2}\prod_{j}\left\{\frac{\sin\frac{a\pi(2j+1)}{k+2}}{\sin\frac{a\pi}{k+2}}\right\}^{s_j}\label{ms1}
\ea
where $a=2q+1$ and the combinatorial factor reflects the statistical distinguishability \cite{qg} and $s_j$ is the number of punctures with spin value $j$,  i.e. $s_{1/2}=$ number of punctures with spin $1/2$, $s_1=$ number of punctures with spin $1$ and so on.

\vspace{0.3cm}
\par
{\it Defining the Microcanonical Ensemble} : A QIH is characterized by the level of the CS theory $(k)$ and the total number of punctures$(N)$. But, it should be noted that in eq.(\ref{ms1}), the basic variables are $s_j$ and $j$. Hence, a microcanonical ensemble of QIHs is defined by fixing $\sA$ and $N$ i.e. the spin configuration $\{s_j\}$ must obey the following constraints
\begin{subeqnarray}\label{con}
&&{\cal C}_1 : \sum^{k/2}_{j=1/2} s_j = N \slabel{con1} \\
&&{\cal C}_2 : \sum^{k/2}_{j=1/2} s_j \sqrt{j(j+1)} = \sA\slabel{con2} 
\end{subeqnarray}
where $\sA=A/8\pi\gamma \lp^2$, $N=$ total number of punctures of the QIH. \footnote{This is analogous to  the microcanonical ensemble of a system of ideal gas where the ensemble is defined by {\it a priori} fixing the total energy $(E=\sum_i\epsilon_in_i)$ and total number of gas particles $(N=\sum_i n_i)$ of the system, where $\epsilon_i$ is the $i$th energy level available for the gas particles. $\epsilon_i$ is the analog of $\j$. Both are known from the underlying quantum theory.}
 Since $A$ involves the Barbero-Immirzi(BI) parameter$(\g)$, whose value is yet to be determined, fixing $A$ will not be a {\it pure} constraint on the variables $s_j$ and $j$. Hence, we take $\sA$ to be a more fundamental entity which involves only the basic variables $s_j$ and $j$. The motivation of considering the above two constraints in eq.(\ref{con}) to be independent comes from the fact that, for large QIH, which we are interested in, arbitrary small fluctuations $\d \sA(\ll \sA)$ and $\d N(\ll N)$ can be considered to be independent, up to a good approximation, assuming the smoothness of the variables.

\vspace{0.3cm}
\par
{\it The Microcanonical Entropy} : The {\it microcanonical} entropy is calculated as follows. Variation of $\log d[\left\{s_j\right\}]$ with respect to $s_j$, subject to the constraints ${\cal C}_1$ and ${\cal C}_2$, yields the dominant configuration which maximizes the entropy of the QIH. The variational equation is given by 
\ba
\d \log d[\left\{s_j\right\}]-\lm\d \sA -\s\d N=0\label{var}
\ea 
where $\d$ represents variation with respect to $s_j$, $\lm$ and $\s$ are the Lagrange multipliers for ${\cal C}_1$ and ${\cal C}_2$ respectively. This yields the dominant configuration given by
\ba
\bar s_j=NM_j(k)e^{-\lm\sqrt{j(j+1)}-\s}\label{dc}					
\ea
where   
\ba
M_j(k)&=&\prod^{k+1}_{a=1}\left\{\f{\sin\f{a\pi(2j+1)}{k+2}}{\sin\f{a\pi}{k+2}}\right\}^{\f{f_a(k)}{f(k)}}\nn\\
f(k)&=&\sum_{a=1}^{k+1}f_a(k)\nn\\
&=&\sum_{a=1}^{k+1}\sin^2\f{a\pi}{k+2}\prod_{j}\left\{\f{\sin\f{a\pi(2j+1)}{k+2}}{\sin\f{a\pi}{k+2}}\right\}^{\bar s_j}\nn
\ea 
Calculating $\log d[\{s_j\}]$ in the limits $\bar s_j,k\to\infty$ for large QIH one can show, following \cite{mpre} (and references therein), that the {\it microcanonical} entropy of a QIH is given by 
 \ba
 S_{MC}=\f{A}{4\lp^2} +N\s(\g) -\f{3}{2}\log \f{A}{\lp^2} \label{smc}
 \ea
where one has to choose $\g=\lm/2\pi$ so as to retrieve the Bekenstein-Hawking area law. This yields $\s(\g)=\log\sum_jM_j(k)e^{-2\pi\g\j}$. Most importantly, using eq.(\ref{dc}) in the eqs.(\ref{con}) along with the identification $\lm=2\pi\g$, one can solve for $\g$ and $\s$ in terms of $\sA/N$. A graphical analysis of the solution of $\g$ plotted as a function of $\sA/N$ reveals that the allowed values of the BI parameter lies in the range given as  $0.159<\g<0.225$ for $4.000<\sA/N<\infty$. This is somewhat different from the formula for the microcanonical entropy of a QIH derived in \cite{GP} where the BI parameter is left as an {\it arbitrary} real positive number.


\section{Energy Spectrum of QIH}\label{es}

In this section we shall discuss about the energy spectrum for the QIH observed by {\it a local observer}, which have been proposed in \cite{GP,GP2} as a model of a black hole.  Since this work is focused towards
studying the thermodynamics of {\it quantum} Isolated Horizon having the energy spectrum proposed in \cite{GP}, we shall be very brief in discussing the aspects of {\it classical} general relativity related to this work (e.g. definition of local observers, approximations of energy expression) to avoid unnecessary lengthening of the paper. Of course we shall point out the proper references where the ideas have been discussed in complete details.

\vspace{0.3cm}
\par
{\it Local observers :} The definition of local observers used in \cite{GP,GP2} originates from the ideas extensively discussed in the textbooks on General Relativity such as \cite{wa}, \cite{fro} etc. These preferred class of observers are {\it stationary} with respect to the horizon of the black hole, which makes them ideal for the observation of the thermodynamics of horizons according to \cite{GP2}. The four velocity of such an observer is given by $u=\f{\xi}{\sqrt{|\xi.\xi|}}$, where $\xi$ is the time-like Killing vector associated with the black hole spacetime and alternatively, the generator of the one-parameter group of isometry  $({\sL}_{\xi}g_{ab}=0)$ for the associated black hole spacetime with the metric $g_{ab}$. While infinitesimally close to the black hole event horizon these observers represent the ZAMOs of \cite{tho}. For more details on this account one may look into \cite{GP2} and the references therein.

\vspace{0.3cm}
\par
{\it Energy observed by a local observer :} The energy spectrum used in \cite{GP} follows from the definition of energy of a classical Schwarzschild black hole from a local observer's perspective. The definition of local energy \cite{GP,GP2} is given by
\ba
E_r = - \frac{1}{8\pi}\int_{S_r}\nabla^a u^b dS_{ab}\label{komar}
\ea
where $u=\f{\xi}{\sqrt{|\xi.\xi|}}$ is the four velocity of the local observer and   $S_r$ is a 2-sphere of radius $r>2M$. For a Schwarzschild black hole, it is straightforward to show that $E_r= \f{M}{2}\left(1-\f{2M}{r}\right)^{-\frac{1}{2}}$. The near horizon limit at $r=2M+\epsilon$ is obtained to be 
\ba
E\equiv E_{2M+\epsilon}\approx\f{M}{2}\left(\f{2M}{\epsilon}\right)^{\f{1}{2}} 
=\frac{A}{8\pi \ell}\label{eal}
\ea
where one has to use $\ell=2(2M\epsilon)^{\frac{1}{2}}$, $A=16\pi M^2$ and consider $\epsilon\ll 2M$.  
The above result has been generalized for the case of Kerr-Newman black hole in \cite{GP2}.

Following the definition of energy given by eq.(\ref{eal}),  the spectrum of the Hamiltonian operator is proposed \cite{GP} to be  given by eq.(\ref{ea}), which in terms of the area spectrum of the QIH in LQG \cite{rovthie} can be written as
\ba
\widehat H|j_1,\cdots, j_N\rangle=\frac{\g \lp^2}{\ell}  \sum_{l} \sqrt{j_l (j_l+1)})\  |j_1,\cdots, j_N\rangle\label{ham}
\ea
where $j_l$ taking values from the set $\{\f{1}{2},1,\f{3}{2},\cdots, \f{k}{2}\}$ is the spin associated with the $l$-th puncture. The length scale $\l$ is the new object introduced, which does not belong to the quantum theory, namely LQG. It denotes the proper distance of a stationary  observer from the event horizon of a Schwarzschild black hole at the radial coordinate $r=2M+\epsilon$ and having an acceleration $1/\l$. Detailed explanation of the derivation of the energy spectrum is available in \cite{GP,GP2}. As far as our work is concerned the crucial point to be noted is that {\it the derivation of the energy spectrum in eq.(\ref{ea}) is dependent on the frame of a class of local observers introduced in \cite{GP2}, characterized by the length scale $\l$. Hence, the energy spectrum of the QIH given by eq.(\ref{ea}) is `A Local Observer's view'.} It follows that the results of the thermodynamic analysis of a QIH with such an energy spectrum, which forms the core matter of this paper,  will be the observation of `a local observer'.  Thus, the title of this paper is justified.

\vspace{0.3cm}
\par
{\it Motivation for the use of the spectrum :} The Hamiltonian formulation of the classical phase space of spacetimes admitting internal boundaries (classical isolated horizons(CIH) at equilibrium) shows that there exists an energy associated with each CIH satisfying a first law\cite{firstlaw}. A correct  quantization of such a theory must lead to a horizon energy spectrum expressed in terms of the spectra of the operators corresponding to the other extensive variables of the first law (namely area, charge, angular momentum, etc.). Unfortunately, such things have not been done up till now. On the other hand, in quantum geometry\cite{qg}, the full Hilbert space of a quantum black hole can be written as $\cal{H}=\cal{H}_V\otimes\cal{H}_S$ modulo some constraints, where $\cal V(\cal S)$ stands for volume (surface). Thus, any generic state $|\Psi\rangle$, of the quantum black hole can be written as
$|\Psi \rangle = |\Psi_{\cal V}\rangle \otimes |\Psi_{\cal S} \rangle$. Hence, any operator which acts on the states of the Hilbert space $\cal H$, say the Hamiltonian $\widehat H$, must have a form \cite{pm1} ${\widehat H} = ({\widehat H}_{\cal V} \otimes \widehat I_{\cal S} + \widehat I_{\cal V} \otimes {\widehat H}_{\cal S} )$ where $\widehat I$ represents identity operator. But the spectrum of this Hamiltonian is unknown. So, clearly there is a missing link between the classical and quantum theories of IH as far as the energy spectrum is concerned. As  already mentioned in \cite{GP}, the effort made in \cite{GP,GP2} is aimed to provide this missing link by an input from the classical theory.

\vspace{0.3cm}
{\it The Canonical Partition function :} Now, using quantum geometry one can show\cite{pm1,ampm} that the partition function of a quantum black hole is completely determined by surface states i.e. $Z=Z_{\cal S}= Tr_{\cal S}\exp-\beta\widehat H_{\cal S}$. This is nothing but the partition function for the QIH. Now, the generic wave function of the QIH can be written as $|\Psi_{\cal S}\rangle=\sum_{j_1,\cdots,j_N}C(j_1,\cdots,j_N)|j_1,\cdots,j_N\rangle$, where $|C(j_1,\cdots,j_N)|^2$ is the probability of finding the QIH in the eigenstate $|j_1,\cdots,j_N\rangle$ having the spin sequence $\{j_1,\cdots,j_N\}$. Hence, following eq.(\ref{eal}), the Hamiltonian acting on the generic wave function of the QIH can be written as \cite{GP}
\ba
\widehat H_{\cal S}|\Psi_{\cal S}\rangle=\frac{1}{8\pi \l}\widehat A|\Psi_{\cal S}\rangle
\ea
where $\widehat A$ is the area operator in LQG \cite{rovthie}. The above equation can be explicitly written in terms of the spectrum of $\widehat H_{\cal S}$ which can be written in terms of the spectrum of $\widehat A$ following eq.(\ref{ea}) as
\ba
\widehat H_{\cal S}|j_1,\cdots, j_N\rangle=\left(\f{\g\lp^2}{\l}  \sum^N_{l=1} \sqrt{j_l (j_l+1)}\right)  |j_1,\cdots, j_N\rangle
\ea
where $\g$ is the Immirzi parameter and $\lp$ is the Planck length. $j_l$ is the spin associated with the $l$ th puncture, $N$ being the total number of punctures. $|j_1,\cdots,j_N\rangle$ is a microstate of the SU(2) CS theory on the QIH, designated by the spin sequence $\{j_1,\cdots,j_N\}$ and also an eigenstate of the Hamiltonian $\widehat H_{\cal S}$. Hence, to be more appropriate $\widehat H\equiv \widehat H_{\cal S}$ in the eq.(\ref{ham}) which has been proposed in \cite{GP}. Also, the trace over the surface states in the partition function discussed above is the sum over  all possible spin sequences.

\vspace{0.3cm}
\par
{\bf Remarks :} The energy spectrum used in \cite{GP} is based on an approximation of the energy, observed by a stationary observer near the horizon of a classical Schwarschild black hole, resulting from some {\it ad hoc} arguments put forward in \cite{GP2}\footnote{ This procedure involves some questionable issues \cite{majhi} and one of which is the subject matter of this paper. Hence, it is not hard to understand that, in spite of the criticism in \cite{majhi}, we have to use the same notion of local energy here for a detailed investigation of its thermodynamic aspects. This is just to clarify in case one wonders why the same author criticizes and uses the same idea.}. 
 Obviously this is not a true quantization of horizon energy. Nevertheless, the thermodynamic aspects of a QIH with the energy spectrum of \cite{GP}, as observed by a local observer, is worth studying as far as its importance in the current literature is concerned, especially \cite{bia} and \cite{smo} besides several other works (not to be listed here). In \cite{bia}, the notion of a quantum Rindler horizon is introduced whose classical version describes the near-horizon geometry of a non-extremal black hole as seen by a stationary local observer. The dynamics of the quantum surface, describing this system, is generated by the boost Hamiltonion of Lorentzian Spinfoams. The crucial point is that the expectation value of this boost Hamiltonian results in the local horizon energy introduced in \cite{GP2}. On the other hand, in \cite{smo} it has been shown that the energy expression given by eq.(\ref{eal}) comes out to be equal to the canonical energy associated with the boundary term of the Holst action, alongside other relevant consequences as far as horizon thermodynamics is concerned. In a nutshell, even though the energy spectrum of a QIH given by eq.(\ref{ea}) is a very specific one, there is much reason to pay attention to it as far as its physical consequences are concerned.

\section{Thermodynamics of QIH}\label{tqih}
This section is dedicated to an exhaustive thermodynamic stability analysis of a QIH having the energy spectrum given by eq.(\ref{ea}), as observed by a local observer.

\subsection{Explicit Quantum Statistical Stability Analysis}\label{eqssa}
To get an insight of the thermodynamic properties of the particular model of quantum black hole, we explore the canonical ensemble scenario where the total number of punctures $(N)$ is kept {\it fixed} and the energy $(E)$ is allowed to fluctuate. The canonical partition function can be written as a sum over spin configurations as
\ba
Z(\b,N)=\sum_{\left\{s_j\right\}}d[\left\{s_j\right\}]e^{-\beta E_{\left\{s_j\right\}}}\approx d[\left\{\bar s_j\right\}]e^{-\beta E} \label{Z}
\ea
where $E_j =$ energy associated with a spin $j$, $~\sum_j\bar s_jE_j= E=$ energy of the QIH for $\{\bar s_j\}$(thermal equilibrium). The contributions from the sub-dominant configurations are neglected.  $\b$ is the inverse temperature of the QIH given by $\b=\p S_{MC}/\p  E|_{N}$ 
which results in
\ba
\b=2\pi\l\left(1-\f{6}{A}\right)\label{beta}
\ea
To calculate relevant thermodynamic quantities one needs to calculate the logarithm of the partition function in the appropriate limits ($\bar s_j,k\to\infty$). A straightforward calculation using equations (\ref{ea}), (\ref{Z}) and (\ref{beta}) yields
\ba
\log Z=  N\s(\g) -\f{3}{2}\log A + \f{3}{2} \label{logz}
\ea

\par

The {\it average energy} of the QIH in the canonical ensemble can be calculated from (\ref{logz}) using the usual thermodynamical relation $\langle E \rangle=-\f{\p}{\p\b} \log Z$.
Using $~d\b/dA=12\pi\l/A^2$ and $E=A/8\pi\l$ it is straightforward to show that the average energy of the QIH is equal to its equilibrium energy i.e. $\langle E \rangle =E$. Following this, the {\it specific heat} of the QIH can be calculated using the usual thermodynamic formula $C=-\b^2\p \langle E \rangle/\p\b$. A few steps of algebra lead to
\ba
C
=-\f{\b^2A^2}{96\pi^2\l^2}\nn
\ea
where one has to use $\langle E \rangle =E=A/8\pi\l$ and $d\b/dA=12\pi\l/A^2$. The specific heat being {\it negative definite} one can conclude that a QIH, having energy spectrum as in eq.(\ref{ea}), is locally \textit{unstable} as a thermodynamic system.

\par
In this context, the validity of {\it the first law} can be checked using (\ref{smc}) and (\ref{temp}) and one can easily show that $dE=TdS$ is indeed satisfied. Also, from (\ref{beta}) one can find the local {\it horizon temperature}  to be 
\ba
 T = \frac{1}{2\pi\l}\left(1+\frac{6}{A}+ .....\right)~~~~~~~~[k_B=1]\label{temp}
\ea
This local horizon temperature contains a series which is identical to the correction terms obtained for the horizon temperature in \cite{ampm} considering Gaussian thermal fluctuations about the equilibrium. The connection between these two may be a future issue of interest.

\par

{\it NOTE :} Let us have a closer look at the canonical partition function. The {\it exact} canonical partition function, without any approximation, can be written as 
\[Z(\b,N)= \bar Z(\b,N) + \d(\b,N)\]
where $\d(\b,N)$ is the contribution from {\it thermal fluctuations} (sub-dominant configurations $\{s_j\}$s other than $\{\bar s_j\}$) about equilibrium value $(\bar Z)$ of the canonical partition function coming from  the dominant configuration $\{\bar s_j\}$ whose spin distribution is given by eq.(\ref{dc}). 
Truly speaking, eq.(\ref{Z}) is only $\bar Z$ and {\it not} $Z$. The effect of the thermal fluctuations is completely neglected ($\d =0$) in eq.(\ref{Z}). This has a profound implication. 

The canonical entropy is given by $S_C= \log Z + \b \langle E\rangle$, which can be recast as
\ba
S_C
&=& \bar S_C + \log (1+\delta/\bar Z) \nn
\ea
where $\bar S_C =\log \bar Z +\b E$ and $\langle E\rangle=E$ is used. If one calculates $\log \bar Z + \b E$, a few steps of algebra leads to $\bar S_C=S_{MC}$ .
Therefore,
\[S_C=S_{MC}+\log (1+\delta/\bar Z)\]
If we do not take the effects of thermal fluctuations in canonical ensemble i.e. $\delta =0$, then it is obvious that $S_C=S_{MC}$ at {\it all} temperatures. This is a very general result concerning a thermodynamic system \cite{landau} which also applies for black holes as has been shown earlier in the literature \cite{therm}. In fact this extra contribution from thermal fluctuations plays a very important role in analyzing the thermodynamic stability of the QIH  \cite{pm1} which we will discuss briefly in the next subsection. Sitting at the equilibrium and ignoring the thermal fluctuations lead to a physically incomplete scenario which apparently looks to give us an {\it ensemble independent} result \cite{GP}. Hence, to get the complete picture, we must take into account quantum and thermal fluctuations both.

\subsection{Thermal Fluctuations in Canonical Ensemble}\label{gtf}
 Starting from the canonical  partition function of a generic QIH and considering Gaussian thermal fluctuations about an equilibrium configuration (saddle point)\cite{landau}, one can actually derive an inequality between the energy and the microcanonical entropy of the QIH at  equilibrium using appropriate units \cite{pm1}. This inequality serves as the stability criterion for the QIH. It is as simple as 
\ba
E>S_{MC}\label{cri}
\ea
where $E$ and $S_{MC}$ are the equilibrium energy and the microcanonical entropy of a generic QIH respectively. Since, here we know both the energy spectrum and the microcanonical entropy of the QIH, it is straightforward to compare eq.(\ref{ea}) and eq.(\ref{smc}) and see that the stability criterion eq.(\ref{cri}) is violated. Hence a QIH having an energy spectrum given by eq.(\ref{ea}) is locally {\it unstable} as a thermodynamic system. To be more explicit, if one calculates the partition function including the Gaussian thermal fluctuations and using $E=A/8\pi\l$, it comes out to be
\ba
Z\approx\f{1}{4\pi}e^{S_{MC}(\bar A)-\b E(\bar A)}\int^{\infty}_0e^{\f{3}{4\bar A^2}a^2}da
\ea
where $\bar A$ is the horizon area at equilibrium(saddle point) and $a$ is the fluctuation variable. The partition function is clearly undefined due to the infinite integral.  Also, if one calculates the canonical entropy($S_C$) taking the Gaussian thermal fluctuations into account\cite{therm}, it comes out to be
\ba
S_C=S_{MC}-\f{1}{2}\log \Delta\nn
\ea
where 
\ba
\Delta=\f{K}{\p E/\p A}\left[\f{\p^2 E}{\p A^2}\f{\p S_{MC}}{\p A}-\f{\p^2 S_{MC}}{\p A^2}\f{\p E}{\p A}\right]
\ea
evaluated at the saddle point(equilibrium configuration), $K$ being an irrelevant positive constant. For the canonical entropy to be well defined we must have $\Delta>0$. But, using the energy spectrum given by (\ref{ea}) it is straightforward to show that $\Delta < 0$. Thus the canonical entropy can not be defined for a QIH having an energy spectrum as in eq.(\ref{ea}) which implies nothing but the instability of the QIH as a thermodynamic system\cite{pm1} from the perspective of a local observer. ($N$ does not play any role as it is kept fixed in canonical ensemble.)  
\par
In this Gaussian approximation method, the inverse temperature of the QIH at equilibrium is given by 
$\b=\f{\p S_{MC}/\p A}{\p E/\p A}$. Using eq.(\ref{ea}) and eq.(\ref{smc}) in the expression for $\b$, it is easy to find that $\b=2\pi\l\left(1-6/A\right)$. Comparing this result with eq.(\ref{beta}), one can see that the equilibrium temperature comes out to be the same in both the approaches. This is a consistency check.

\vspace{0.3cm}
\par
{\bf Remarks :} The alternative {\it thermal holographic} approach for the thermodynamic stability analysis of a system, shown immediately above, was first adopted in the context of black hole thermodynamics in canonical ensemble in \cite{pm1} and generalized to the case of grand canonical ensemble in \cite{ampm}, following the elegant textbook of statistical physics by Landau and Lifschitz \cite{landau}. The positivity of the coefficient of the second order term in the exponent of the integral of the partition function is equivalent to the negativity of the specific heat of the system under consideration (e.g. see \cite{monteiro} for an explicit deduction). This method gives us an explicit visualization of the equilibrium and the fluctuation contributions to the thermodynamic quantities, thus providing a more physical insight to the problem under investigation. Moreover, this simple and elegant method provides us with a very fundamental criterion for the stability of quantum black holes given by the inequality (\ref{cri}) resulting from a theory {\it devoid of the use of any classical metric}. Given the energy of a black hole (microcanonical entropy being known from the quantum theory), the inequality (\ref{cri}) serves as a ``testing tool'' for the examination of the thermodynamic stability of that particular black hole. In fact, one can check that the inequality (\ref{cri}) successfully explains the instability of Schwarzschild, Reissner-Nordstrom black holes and the stability of AdS-Schwarzschild and AdS-Reissner-Nordstrom black holes \cite{pm1,ampm}.

\section{Discussion}\label{d}
Let us conclude with a brief summary of the thermodynamic stability analysis of the particular model of a QIH presented in this paper. The final conclusion is the local thermodynamic \textit{instability} of a QIH having energy spectrum given by the eq.(\ref{ea}). The result has been derived in two different approaches shown in the subsections (\ref{eqssa}) and (\ref{gtf}). The crucial role of thermal fluctuations behind the thermodynamic instability of this particular model of the QIH is evident from the analyses. In fact, though the quantum statistical analysis gives a negative specific heat, the physical picture becomes much clearer in the {\it thermal holographic} analysis where the Gaussian thermal fluctuations manifestly control the convergence criterion of the partition function. In this approach we can actually see that any arbitrary QIH is {\it not} thermodynamically unstable, but only those which fail to satisfy the convergence condition for the partition function are unstable. The particular energy spectrum of the QIH given by eq.(\ref{ea}) considered in this paper is only one such example. This is {\it not} the stability analysis of a generic QIH, for which we need the information about the quantum energy spectrum of a QIH derived from a true quantization resulting from the fundamental quantum theory. It can be only considered to be a stability analysis of the QIH from the perspective of a local observer.

\par
Now, an alert reader will surely wonder : {\it Why shall we make a thermodynamic analysis of such a very specific energy spectrum of a QIH ?} The answer is very simple : {\it The particular energy spectrum studied in this paper is of utmost importance as far as current literature is concerned.} Some of the important aspects of the particular definition of energy of a QIH studied in this work  has been mentioned in the ``Remarks'' at end of section(\ref{es}) in connection with current literature. But the most important consequence has been reported in \cite{GP}. Using the energy spectrum in eq.(\ref{ea}), it has been claimed in \cite{GP} that {\it `` as a thermodynamic system the Isolated Horizon is locally stable''}.
Hence, the purpose of an extensive thermodynamic analysis of a QIH having the particular energy spectrum given by eq.(\ref{ea}) is now clear enough and it does not need much of an effort to understand that our results are in complete contradiction with the above claim.

\vspace{0.5cm}
\textbf{Acknowledgments :} I gratefully acknowledge Prof. Parthasarathi Majumdar for his encouragement and advice besides the long illuminating discussions regarding the important issues of this work.



\end{document}